\documentclass{jetpl}
\twocolumn

\DeclareMathOperator{\Tr}{Tr}
\DeclareMathOperator{\tr}{tr}
\DeclareMathOperator{\intern}{int}
\DeclareMathOperator{\diag}{diag}

\DeclareMathOperator{\eff}{eff}
\DeclareMathOperator{\cntr}{}
\DeclareMathOperator{\QM}{QM}
\DeclareMathOperator{\ECL}{ECL}
\DeclareMathOperator{\SE}{SE}

\DeclareMathOperator{\gh}{gh}
\DeclareMathOperator{\gf}{g.f.}
\DeclareMathOperator{\const}{const}
\DeclareMathOperator{\MM}{(M)}
\DeclareMathOperator{\OM}{(0,M)}
\DeclareMathOperator{\OMM}{(0,MM)}
\DeclareMathOperator{\GG}{\mathfrak{G}}
\newcommand{\ket}[1]{|#1\rangle}
\newcommand{\bra}[1]{\langle #1|}
\newcommand{\be}{\begin{equation}}
\newcommand{\ee}{\end{equation}}
\newcommand{\bea}{\begin{equation}\begin{aligned}}
\newcommand{\eea}{\end{aligned}\end{equation}}

\newcounter{mysect}
\newcommand{\newsect}{\par\vspace{0.5 cm}\addtocounter{mysect}{1} \noindent {\bf \arabic{mysect}.} }

\title{Pseudoscalar mesons and their radial excitations from the Effective Chiral Lagrangian}

\rtitle{Pseudoscalar mesons and their radial excitations \ldots}

\sodtitle{Pseudoscalar mesons and their radial excitations from the Effective Chiral Lagrangian}

\author{S.\,M.\,Fedorov\/\thanks{fedorov@heron.itep.ru},
        Yu.\,A.\,Simonov\/\thanks{simonov@heron.itep.ru} }

\rauthor{S.\,M.\,Fedorov, Yu.\,A.\,Simonov}

\sodauthor{Fedorov~S.\,M., Simonov~Yu.\,A.}

\address{Institute of Theoretical and Experimental Physics,
117218, Moscow, B.Cheremushkinskaya 25, Russia}

\dates{June 23, 2003}{*} %!!!

\abstract{
Effective Chiral Lagrangian is derived from QCD in the framework of Field
Correlator Method. It contains the effects of both confinement and
chiral symmetry breaking due to a special structure of the resulting quark mass operator. It is
shown that this Lagrangian describes light pseudoscalar mesons, and
Gell-Mann-Oakes-Renner relations for pions, eta and K mesons are
reproduced. Spectrum of radial excitations of pions and K mesons is found and compared to
experimentally known masses.
}

\begin{document}

\maketitle

\newsect
QCD is known to possess two highly nontrivial features at low temperatures, namely confinement and
chiral symmetry breaking (CSB). At some critical temperature phase transitions of deconfinement
and chiral symmetry restoration occur. From lattice calculations it is known that these two phase transitions
take place at the same temperature~\cite{Karsch:1998ua,Carmona:2001tn}. The fact that two critical
temperatures coincide was not fully understood so far. This work is a continuation of a series
of papers~\cite{Simonov:2002er,Simonov:2003gx,Simonov:2003kx}, where it is argued that CSB occurs
due to confinement in a very nontrivial way.

It was shown in~\cite{Simonov:2002er} that effective four-quark interaction leading to spontaneous
chiral symmetry breaking, occurs in QCD due to confinement, and is associated with the QCD string.
Thus, CSB is closely connected to confinement. In this approach the Effective Chiral Lagrangian (ECL)
containing fields of light pseudo-scalar mesons is derived from QCD Lagrangian. This is done by
integrating out gluon fields and performing bosonisation. At the same time confinement is
taken into account through specific form of gluon-field correlators.

As a result, expanding in powers of (derivatives) of bosonic fields, one obtains the ECL
similar to the celebrated Gasser-Leutwyler Lagrangian~\cite{Gasser:1983yg}, however in the nonlocal
form~\cite{Simonov:2002er}.

We expand ECL in powers of meson fields, and reproduce standard Gell-Mann-Oakes-Renner relations, while meson masses
are zero in the chiral limit. It is shown that the vanishing of meson masses happens due to cancellation of two terms in
Green's functions of mesons. Poles of Green's function corresponding to radial excitations of pseudoscalar mesons
are displaced from the masses, obtained in Hamiltonian approach without CSB effects (see e.g.~\cite{Simonov:1999qj} and
references therein), and are shifted down by less than 15 \%.

\newsect
We consider Euclidean partition function for quarks and gluons in the presence of external classical
currents $v_{\mu}$, $a_{\mu}$, $s$ and $p$
\bea
Z &= \int DA D\bar\psi D\psi \exp \left[ - (S_0 + S_1 + S_{\intern} + S_{\gf} + S_{\gh})\right],\\
S_0 &= \frac{1}{4} \int d^4 x \left( F_{\mu\nu}^a\right)^2,\\
S_1 &= -i \int d^4 x \bar \psi^f (\hat \partial + \hat v + \gamma_5 \hat a + s + i\gamma_5 p)^{fg}\psi^g,\\
S_{\intern} &= - \int d^4 x \bar \psi^f g \hat A^a t^a \psi^f.
\eea
Here $f,g=1,2,3$ are flavor indices, $t^a$ are generators of color SU(3) group, $\tr t^a t^b = \delta^{ab}/2$,
$a=1,\ldots 8$. $S_{\gf}$ and $S_{\gh}$ are gauge fixing and ghost terms.

Next, we use the generalized contour gauge~\cite{Ivanov,Shevchenko:1998uw}
\be
A_{\mu}(x) = \int_0^1 ds \frac{\partial z_{\nu}(s,x)}{\partial s} \frac{\partial z_{\rho}(s,x)}{\partial x_{\mu}}
             F_{\nu\rho}(z(s)).
\ee
Here $z_{\nu}(s,x)$ belongs to a set of contours, with properties: $z_{\nu}(0,x)=x_0$,
$z_{\nu}(1,x)=x_{\nu}$, $x_0$ is a fixed point.
In what follows the exact position of contours is unimportant for our analytical results, while
for numerical estimates we will assume that contours are chosen to minimize the spectrum (and
area of the string world sheet), to be called the minimal set of contours.

The reason we use contour gauge is that it allows to express gauge field $A_{\mu}$ through field
strength tensor $F_{\mu\nu}$. Now we are in position to integrate out gluon field $A_{\mu}$, expressing
the result in terms of field correlators:
\be
\begin{aligned}
&Z = \int \int D\bar\psi D\psi \exp \left[ - (S_1 + S_{\eff} )\right],\\
&\exp\left[ - S_{\eff} \right] = \langle \exp \left[ - S_{\intern}\right] \rangle_{A}
\end{aligned}
\ee
We use cluster expansion to evaluate this average over gluon fields
\be
\label{eq_clstr}
\begin{aligned}
&\langle \exp \left[ - S_{\intern}\right] \rangle_{A} = \exp \left( \sum_n \frac{(-1)^n
    \langle\!\langle S_{\intern}^n \rangle\!\rangle}{n!} \right), \\
&\langle\!\langle S_{\intern} \rangle\!\rangle = \langle S_{\intern} \rangle_{A}\equiv 0,\\
&\langle\!\langle S_{\intern}^2 \rangle\!\rangle = \langle S_{\intern}^2 \rangle_{A} - \langle S_{\intern} \rangle_{A} ^2 = \langle S_{\intern}^2 \rangle_{A},\\
&\ldots
\end{aligned}
\ee
It is clear that gauge invariant quantities like spectrum and Green's functions computed
with the help of $S_{\eff}$ do not depend on the chosen contours, if all terms of cluster
expansion are retained in~(\ref{eq_clstr}). In what follows we will
use the Gaussian approximation, and consider only first
two terms in cluster expansion, $n=1,2$. As was shown in~\cite{Shevchenko:2000du,Dosch:2000va} the Gaussian
approximation on minimal surfaces is accurate within few percent. Thus, we have
\be
\begin{aligned}
& S_{\eff} = -\frac{1}{2} \int d^4 x d^4 y
  \bar \psi^f_{i\alpha} (x) \psi^f_{j\beta}(x) \bar \psi^g_{k\gamma}(y) \psi^g_{l\delta}(y) \times\\
&\times
  \int_0^1 ds dt \frac{\partial z_{\rho}(s,x)}{\partial s} \frac{\partial z_{\lambda}(s,x)}{\partial x_{\mu}}
  \frac{\partial z_{\rho'}(t,y)}{\partial t} \frac{\partial z_{\lambda'}(t,y)}{\partial y_{\nu}} \times\\
&\times  \left\langle \left[F_{\rho\lambda}(z(s,x))\right]_{ij} \left[F_{\rho'\lambda'}(z(t,y))\right]_{kl}\right\rangle_A
  \left(\gamma^{\mu}\right)_{\alpha\beta} \left(\gamma^{\nu}\right)_{\gamma\delta}.
\end{aligned}
\ee
Here $i$,$j$,$k$,$l$ are color indices, $\alpha$,$\beta$,$\gamma$,$\delta$ are spinor indices.
Inserting parallel transporters $\Phi(x,x_0)$ and $\Phi(y,x_0)$, which are identically equal to
unity in contour gauge, one finally gets expression for
effective action
\be
\label{eq_seff}
\begin{aligned}
&S_{\eff}=
-\frac{1}{2} \int d^4 x d^4 y
  \bar \psi^f_{i\alpha} (x) \psi^f_{j\beta}(x) \bar \psi^g_{k\gamma}(y) \psi^g_{l\delta}(y)
  \times\\
  &\quad \quad \times \left( \delta_{jk}\delta_{il} - \frac{1}{N_c}\delta_{ij}\delta_{kl} \right)
  J_{\alpha\beta\,\gamma\delta}^{\cntr}(x,y),\\
&J_{\alpha\beta\,\gamma\delta}^{\cntr}(x,y)=
  \left(\gamma_{\mu}\right)_{\alpha\beta} \left(\gamma_{\nu}\right)_{\gamma\delta} J_{\mu\nu}^{\cntr}(x,y),\\
&J_{\mu\nu}^{\cntr}(x,y)= \frac{1}{N_c^2-1} \times \\
  & \quad \times \int_0^1 ds dt \frac{\partial z_{\rho}(s,x)}{\partial s} \frac{\partial z_{\lambda}(s,x)}{\partial x_{\mu}}
    \frac{\partial z_{\rho'}(t,y)}{\partial t} \frac{\partial z_{\lambda'}(t,y)}{\partial y_{\nu}} \times\\
  & \quad \times \tr \left\langle F_{\rho\lambda}\left(z(s,x),x_0\right) F_{\rho'\lambda'}\left(z(t,y),x_0\right)\right\rangle_A, \\
&F(u,x_0) \equiv \Phi(x_0,u) F(u) \Phi(u,x_0).
\end{aligned}
\ee
Performing bosonization, and keeping only scalar-isoscalar and pseudoscalar-isovector (corresponding to
pions, $K$ and $\eta$ mesons) terms, one arrives at the
quark-meson Lagrangian (see~\cite{Simonov:2003gx} for details):
\be
\begin{aligned}
&Z = \int D\bar\psi D\psi D M_s D \phi_a \exp \left[ - S_{\QM}\right],\\
&S_{\QM} = - \int d^4 x d^4 y \Bigl[ \bar \psi^f_{i \alpha} (x) \times\\
  &\quad\times\Bigl( i (\hat \partial + \hat v + \gamma_5 \hat a + s + i\gamma_5 p)^{fg}_{\alpha\beta} \delta^{(4)}(x-y) + \\
&\quad\quad + i M_s (x,y) \hat U_{\alpha \beta}^{fg}(x,y)\Bigr) \psi^g_{i \beta} (y) - \\
&\quad\quad\quad - 2 N_f \left( J(x,y)^{\cntr}\right)^{-1} M_s^2(x,y) \Bigr],\\
&J^{\cntr}(x,y) = J_{\mu\mu}^{\cntr}(x,y),\\
&\hat U_{\alpha \beta}^{fg}(x,y) = \exp\left(i \gamma_5 t_a \phi_a (x,y) \right)_{\alpha \beta}^{fg}.
\end{aligned}
\ee
It is now straightforward to integrate out quark fields to obtain the Effective Chiral Lagrangian:
\be
\label{eq_ECL}
\begin{aligned}
&Z = \int D M_s D\phi_a \exp \left[ - S_{\ECL}^{\cntr}\right],\\
&S_{\ECL}^{\cntr}=2N_f\int d^4 x d^4 y \left( J^{\cntr}(x,y)\right)^{-1} M_s^2(x,y) + W(\phi),\\
&W(\phi) = N_c \tr \ln\Bigl[ i (\hat \partial + \hat v + \gamma_5 \hat a + s + i\gamma_5 p) + \\
&\quad +  i M_s(x,y)e^{i \gamma_5 t_a \phi_a(x,y)}\Bigr].
\end{aligned}
\ee
Here tr refers to flavor and spinor indices and to space coordinates. $M_s$ is the effective quark mass operator,
and $\phi_a$ are fields of pseudoscalar mesons (up to the dimensional factor $2/f$, $f$ is the decay constant,
$\phi_a = 2 \pi_a /f$).

Classical equations of motion are given by
\be
\begin{aligned}
&N_c \Tr \left(
  - S_{\phi}(x,y) M_s(x,y) e^{i \gamma_5 t_a \phi_a(x,y)} \gamma_5 t_a \right) = 0,\\
&N_c \Tr \left(
 i S_{\phi}(x,y) e^{i \gamma_5 t_a \phi_a(x,y)}\right) - \\
& \quad - 4 N_f \left(J^{\cntr}(x,y)\right)^{-1}M_s(x,y) = 0, \\
&S_{\phi}(x,y) \equiv
  \bra{x} \frac{1}{i \hat \partial +
    i M_s e^{i \gamma_5 t_a \phi_a} } \ket{y}
\end{aligned}
\ee
This leads to solutions
\be
\label{eq_Ms0_phi0}
\begin{aligned}
&\phi_a^{(0)}(x,y)=0,\\
&M_s^{(0)}(x,y)=\frac{N_c}{4 N_f} J^{\cntr}(x,y) \Tr \left( S(x,y) \right),\\
&S(x,y) \equiv S_{\phi}(x,y)\bigr|_{\phi=0}.
\end{aligned}
\ee
Second equation in~(\ref{eq_Ms0_phi0}) is a nonlinear equation for $M_s^{(0)}$, and the existence
of a nontrivial solution is a manifestation of the chiral symmetry breaking, since $M_s^{(0)}$ is
scalar. The system of equations~(\ref{eq_Ms0_phi0}) for $M_s^{(0)}$ and $S(x,y)$ was considered
in~\cite{Simonov:1997bh} for the special case of heavy-light mesons, and it was shown that it has a confining
scalar solution for $M_s^{(0)}(x,y) \simeq \sigma |\vec x - \vec x_0| \delta^{(3)}(x-y)$ for
large distance $|\vec x - \vec x_0|$ from quark to (heavy) antiquark at the point $\vec x_0$. It
is clear that the same type of solution occurs at large interquark distances for light-light mesons,
which means that confinement and CSB occur spontaneously and simultaneously from the nontrivial solution of
the system~(\ref{eq_Ms0_phi0}).

\newsect
We consider ECL~(\ref{eq_ECL}), expanding it in powers of the field $\phi_a$ up
to the second order, and we introduce current quark masses ${\cal M}_f \equiv  \diag(m_u,m_d,m_s)$.
Neglecting external currents, one obtains
\be
\label{eq_W2}
\begin{aligned}
&W(\phi) = N_c \tr \ln\left[ i (\hat \partial + {\cal M}_f + M_s e^{i \gamma_5 t_a \phi_a})\right] =\\
&\begin{aligned}
    = & N_c \tr \ln\Bigl[ i (\hat \partial + {\cal M}_f + M_s ) + \\
      & \quad + \left(-M_s\gamma_5 t_a \phi_a - \frac{i}{2} M_s t_a t_b \phi_a\phi_b\right) \Bigr],
   \end{aligned}\\
&W^{(2)}(\phi)= - \frac{N_c}{2} \tr \Bigl[i S \cdot \left(M_s t_a t_b \phi_a \phi_b\right) + \\
&\quad\quad +
  S \cdot \left(M_s t_a \phi_a\right) \cdot \gamma_5 S \gamma_5 \cdot \left(M_s t_b \phi_b\right)
  \Bigr].\\
\end{aligned}
\ee
Taking trace in flavor indices, (see~\cite{Simonov:2003gx} for details), one arrives
at the following expression for the term quadratic in meson fields:
\be
\label{eq_W2_Wpipi_etc}
\begin{aligned}
&W^{(2)}(\phi) = -\int d^4 x d^4 y
\Bigl[
  W_{\pi\pi}(x,y) \phi_{\pi}^*(x) \phi_{\pi}(y) + \\
& \quad + W_{KK}(x,y) \phi_{K}^*(x) \phi_{K}(y) + \\
& \quad + W_{K^0 K^0}(x,y) \phi_{K^0}^*(x) \phi_{K^0}(y) + \\
& \quad + \frac{1}{2} \sum_{i,j=3,8} W_{ij}(x,y) \phi_i(x) \phi_j(y)
\Bigr],
\end{aligned}
\ee
where, for example,
\be
\label{eq_ex_Wpipi_etc}
\begin{aligned}
&W_{\pi\pi}(x,y) = \frac{N_c}{4} \Tr \Bigl[
S_u(x,y) M_s(y) \gamma_5 S_d(y,x) \gamma_5 M_s(x) + \\
&\quad + i S_u(x,x) M_s(x) \delta^4(x-y) + (u \leftrightarrow d)  \Bigr].
\end{aligned}
\ee
$S_u$, $S_d$ and $S_s$ are quark propagators~(\ref{eq_Ms0_phi0}) with current mass of corresponding quark
in the denominator.
Here we have taken the local limit of nonlocal operators $M_s(x,y)\to M_s(x)\delta^4(x-y)$,
$\phi(x,y)\to \phi(x)$, which is obtained when gluonic correlation length $T_g$ in the correlator
$\langle F F\rangle$ tends to zero.

Two terms in~(\ref{eq_ex_Wpipi_etc}) correspond to connected and disconnected diagrams, which cancel each other
in the zero momentum limit. This cancellation is exact in the chiral limit. To be more precise, quadratic
term for zero momentum (i.e. when $\phi(x)=\const$) takes the form:
\be
\label{eq_W2_zero_k}
\begin{aligned}
&W^{(2)}(\phi)\bigr|_{\mbox{\footnotesize{zero momentum}}} = \\
&= \frac{N_c}{4} \int d^4 x \Bigl[ \frac{m_u+m_d}{2} \Tr \left( -i S_u(x,x) -i S_d(x,x) \right) \phi_{\pi}^* \phi_{\pi} + \\
&\quad\quad\quad + \ldots \Bigr] + O(m^2).
\end{aligned}
\ee
Taking into account that
\be
\label{eq_bar_psi_psi}
\langle \bar \psi \psi \rangle_M = - i \langle \bar \psi \psi \rangle_E = - \frac{1}{Z} \frac{\delta Z[v,a,s,p]}{\delta s(x)}
  = N_c \Tr \left( i S(x,x) \right),
\ee
where $\langle \bar \psi \psi \rangle_M$ and $\langle \bar \psi \psi \rangle_E$ denote quark condensate in Minkovski
and Euclidean space respectively, and that $\phi_a = 2 \pi_a /f$, $f$ is the decay constant, $\pi_a$ are physical
meson fields, one finds
\be
\label{eq_GMOR}
\begin{aligned}
&f^2 M_{\pi^{\pm}}^2 = 2 \hat m \left| \langle \bar q q \rangle \right| + O(m^2)\\
&f^2 M_{\pi^0}^2 = 2 \hat m \left| \langle \bar q q \rangle \right|  - \varepsilon + O(\varepsilon^2) + O(m^2)\\
&f^2 M_{K^{\pm}}^2 = (m_u + m_s) \left| \langle \bar q q \rangle \right| + O(m^2) \\
&f^2 M_{K^{0}}^2 = (m_d + m_s)\left| \langle \bar q q \rangle \right| + O(m^2)\\
&f^2 M_{\eta_{8}}^2 = \frac{2}{3} (\hat m + 2 m_s) \left| \langle \bar q q \rangle \right| +
  \varepsilon + O(\varepsilon^2) + O(m^2).
\end{aligned}
\ee
Here $\hat m = (m_u + m_d)/2$. We have neglected differences between quark condensates for different
flavors, corrections are of order of $m_q^2$. Small mixing of $\phi_3$ and
$\phi_8$ states due to isospin symmetry breaking (proportional to $m_u-m_d$) yields a correction $\varepsilon$
to pion and $\eta$ meson masses:
\be
\begin{aligned}
&\pi^0 \sim \cos (\delta) \phi_3 + \sin(\delta) \phi_8, \\
&\eta_8 \sim -\sin(\delta) \phi_3 + \cos(\delta) \phi_8, \\
&\tan(2\delta) = \sqrt{3} \frac{m_d-m_u}{2 m_s - (m_u+m_d)},\\
&\varepsilon = \frac{\left| \langle \bar q q \rangle \right| (m_u - m_d)^2}{4 m_s - 2 (m_u + m_d)},\\
&\delta \simeq 0.6^{\circ}.
\end{aligned}
\ee
Thus, ECL~(\ref{eq_ECL}) leads to correct Gell-Mann-Oakes-Renner relations for all light
pseudoscalar mesons.

\newsect
Let us now consider Green's functions of mesons, generated by the pseudoscalar currents:
\be
\begin{aligned}
&\GG_{ab}(x,y) = \langle J_a^5(x) J_b^5(y) \rangle = \frac{1}{Z} \frac{\delta^2 Z}{\delta p_a(x) \delta p_b(y)},\\
&J_a^5(x) = \bar \psi(x)\gamma_5 t_a \psi(x),\\
&p^{fg}(x) = p_a(x) t_a^{fg}.
\end{aligned}
\ee
From the ECL one obtains
\bea
\label{eq_Gab}
&\GG_{ab}(x,y) = \frac{1}{Z} \int \ DM_s D\phi_a \exp \left[-S_{\ECL}\right] \times\\
&\times \Bigl [ N_c \Tr \left(S_{\phi}(x,y) \gamma_5 t_a S_{\phi}(y,x) \gamma_5 t_b \right) - \\
&\quad - N_c^2 \Tr \left(S_{\phi}(x,x) \gamma_5 t_a \right) \Tr \left(S_{\phi}(y,y) \gamma_5 t_b \right) \Bigr]
\eea
Taking $M_s$ at the stationary point~(\ref{eq_Ms0_phi0}) and expanding $S_{\phi}$ in terms of $\phi$
around $\phi_a^{(0)}=0$, one finds:
\be
S_{\phi}(x,y) = S(x,y) + \int d^4 z S(x,z) \cdot M_s(z)\gamma_5\phi_a(z)t_a \cdot S(z,y)
\ee
As argued in~\cite{Simonov:2003gx}, the coupling constant $g_{\pi q \bar q}$ is of the order of
$N_c^{-1/2}$, and thus in large $N_c$
limit pion exchanges are suppressed. This allows to neglect pion fields in connected terms (first term
in equation~(\ref{eq_Gab})), and consider only one pion exchange in the disconnected term. Resulting expression will
contain two terms, both of order of $N_c$.

Taking into account, that $S(x,y) = \diag(S_u(x,y),S_d(x,y),S_s(x,y))$ is diagonal in flavor, one finds:
\be
\begin{aligned}
&\GG_{\pi^+\pi^+}(x,y) = \frac{N_c}{2} \Tr \left( S_d(x,y) \gamma_5 S_u(y,x) \gamma_5 \right) - \\
& - \frac{N_c^2}{4} \int d^4 z_1 d^4 z_2
    \Tr \left( S_u(x,z_1) M_s(z_1) \gamma_5 S_d(z_1,x) \gamma_5 \right) \times\\
&\quad \times   \Tr \left( S_d(y,z_2) M_s(z_2) \gamma_5 S_u(z_2,y) \gamma_5 \right)
                 G^{\phi}_{\pi\pi} (z_1,z_2).
\end{aligned}
\ee
Other Green functions differ only in flavor indices.
Here $G^{\phi}_{\pi\pi}(z_1,z_2) = \langle \phi_{\pi}^*(z_1) \phi_{\pi}(z_2) \rangle$ is
the propagator of pion field.
This formula can be illustrated with the Feynman diagram:
\bea
&\GG_{\pi^+\pi^+}(x,y) = \frac{N_c}{2} \; \parbox{20mm}{\includegraphics{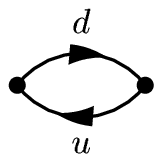}} - \\
& \quad\quad - \frac{N_c^2}{4} \;\parbox{55mm}{\includegraphics{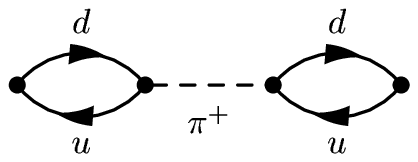}}
\eea

To find the pion propagator one should consider equations~(\ref{eq_W2_Wpipi_etc},\ref{eq_ex_Wpipi_etc}):
\be
\label{eq_Gphi_pipi}
\begin{aligned}
&\left(G^{\phi}_{\pi\pi} (x,y)\right)^{-1} = W_{\pi\pi}(x,y) = \\
&=\frac{N_c}{4}
  \Tr \Bigl[ 2 S_d(x,y) M_s(y) \gamma_5 S_u(y,x) M_s(x) \gamma_5  + \\
& + i \bigl( S_d(x,x) M_s(x) + S_u(y,y) M_s(y) \bigr) \delta^4(x-y)  \Bigr]
\end{aligned}
\ee
Going over to the momentum space $\GG(x,y) = \int d^4 k /(2 \pi)^4 \exp(ik(x-y)) \GG(k)$, where we have also
taken into account that Green function is translational invariant, i.e. depends only on $(x-y)$, one has:
\be
\begin{aligned}
&\GG_{\pi^+\pi^+}(k) =  \frac{N_c}{2} G_{\pi^+\pi^+}^{(0)}(k)  - \\
&\quad -  \frac{N_c^2}{4} G_{\pi^+\pi^+}^{\OM}(k) G^{\phi}_{\pi^+ \pi^+}(k) G_{\pi^+\pi^+}^{\OM}(k),\\
&G_{\pi^+\pi^+}^{(0)}(x,y) \equiv  \Tr \left( S_d(x,y) \gamma_5 S_u(y,x) \gamma_5 \right),\\
&G_{\pi^+\pi^+}^{\OM}(x,y) \equiv  \Tr \left( S_u(x,y) M_s(y) \gamma_5 S_d(y,x) \gamma_5 \right).
\end{aligned}
\ee
Due to Eqs.~(\ref{eq_W2_zero_k} -- \ref{eq_GMOR}) the pion propagator~(\ref{eq_Gphi_pipi}) has pole
at $k^2=-M_{\pi^{\pm}}^2$ and can be rewritten as
\bea
\label{eq_Gpipi_k}
&G^{\phi}_{\pi^+ \pi^+}(k) = \frac{2}{N_c}
    \frac{1}{G_{\pi^+\pi^+}^{\OMM}(k) - G_{\pi^+\pi^+}^{\OMM}(k^2 = -M_{\pi^{\pm}}^2)}, \\
&G_{\pi^+\pi^+}^{\OMM}(x,y) \equiv  \Tr \left( S_u(x,y) M_s(y) \gamma_5 S_d(y,x) M_s(x) \gamma_5 \right).
\eea
As argued in~\cite{Simonov:2003gx}, all three Green functions $G_{\pi^+\pi^+}^{(0)}$,
$G_{\pi^+\pi^+}^{\OM}$, and $G_{\pi^+\pi^+}^{\OMM}$ have the same set of poles, which
are poles of the quark model (i.e. confined $\bar q q$ system without chiral symmetry breaking) in pseudo-scalar
channel, and can be represented as
\bea
&G^{(0)}_{\pi^+\pi^+}(k) = - \sum_{n=0}^{\infty} \frac{c_n^2}{k^2+m_n^2},\\
&G^{\OM}_{\pi^+\pi^+}(k) = - \sum_{n=0}^{\infty} \frac{c_n c_n^{\MM}}{k^2+m_n^2}, \\
&G^{\OMM}_{\pi^+\pi^+}(k) = - \sum_{n=0}^{\infty} \frac{\left(c_n^{\MM} \right)^2}{k^2+m_n^2},
\eea
where
\be
\begin{aligned}
&c_n = \sqrt{\frac{m_n}{2}} \varphi_n(0),\\
&c_n^{\MM} = \sqrt{\frac{m_n}{2}} M(0) \varphi_n(0),
\end{aligned}
\ee
$\varphi_n({\bf r})$ is the 3D spin-singlet wave function of $\bar q q$ system, and $M(0)$ is a constant related to
mass operator $M_s$, evaluated in~\cite{Simonov:2003kx} through $\sigma=0.18$~GeV$^2$ and $T_g=1$~GeV$^{-1}$ to
be $M(0) = 148$~MeV. Thus one has for the pion Green function:
\bea
\label{eq_G_pipi}
&\GG_{\pi^+\pi^+}(k) = -\frac{N_c}{2} \frac{\Psi(k)}{(k^2 + M_{\pi^{\pm}}^2) \Phi(k)},\\
&\Psi(k) = \sum_{n,m = 0}^{\infty} \frac{c_n^2 \left(c_m^{\MM}\right)^2}{(k^2+m_n^2)(m_m^2-M_{\pi}^2)},\\
&\Phi(k) = \sum_{n=0}^{\infty} \frac{\left(c_n^{\MM}\right)^2}{(k^2+m_n^2)(m_n^2-M_{\pi}^2)}.
\eea
Clearly, the Green function~(\ref{eq_G_pipi}) has pole at $k^2 = - M_{\pi^{\pm}}^2$, and all poles of quark
model are cancelled, since the same set of poles appears in functions $\Psi(k)$ and $\Phi(k)$.
The radial excitations of $\pi^{\pm}$ meson are given by zeros of the function $\Phi(k)$. In
the first approximation it reads:
\bea
\label{eq_delta1}
&k^2 = - m_1^2 ( 1 + \delta_1),\\
&\delta_1 = - \frac{1}{m_1^2}
            \frac{c_1^2 (m_1^2-m_0^2)(m_0^2-M_{\pi^{\pm}}^2)}
                 {c_1^2 (m_0^2 - M_{\pi^{\pm}}^2) + c_0^2 (m_1^2 - M_{\pi^{\pm}}^2)}
\eea
Masses of $K^0$, $\bar K^0$ radial excitations can be found from~(\ref{eq_delta1}) with the exchange
of $\pi$ meson mass and reference spectrum with those for $K$ mesons.
Numerical results for masses
of radial excited states are presented in the next section.

It should be mentioned, that $\eta$ meson requires separate consideration, because of it's mixing
with isoscalar state $\eta'$, which is different for mesons and their radial excitations.
Study of $\eta$ meson spectrum and mixings is planned for the future work.

\newsect
Masses and wave functions of reference spectrum can be obtained from the QCD string Hamiltonian
(first derived in~\cite{Simonov:1989fd,Dubin:vn,Dubin:1993fk}, and improved to take into account quark
self-energy in~\cite{Simonov:2001iv}), where we have put $L=0$:
\be
\label{eq_Hamiltonian}
H = \frac{m_1^2}{2 \mu_1} + \frac{m_2^2}{2 \mu_2} + \frac{\mu_1 + \mu_2}{2} + \frac{p_r^2}{2 \tilde \mu}
+ \sigma r - \frac{4}{3} \frac{\alpha_s}{r}.
\ee
Here $m_1$ and $m_2$ are current masses of quarks, $\mu_1$ and $\mu_2$ are einbein parameters, to be found from the
eigenvalues of Hamiltonian~(\ref{eq_Hamiltonian}) via  $\partial \bar M_n(\mu_1,\mu_2) / \partial \mu_1 = 0$,
$\partial \bar M_n(\mu_1,\mu_2) / \partial \mu_2 = 0$; $\tilde \mu = \mu_1 \mu_2 /(\mu_1 + \mu_2)$, and $p_r$ is the
radial component of momentum. This Hamiltonian allows to find spin averaged masses and wave functions.
Spin-spin interaction can than be taken into account as a perturbation:

\bea
&M_n = \bar M_n (\mu_1,\mu_2) + \frac{32 \pi \alpha_s \vec{s_1}\vec{s_2}}{9 \mu_1 \mu_2}|\varphi_n(0)|^2 + \\
&\quad +
\frac{4}{3}  \langle \frac{\alpha_s}{r^3}\rangle
  \frac{\langle 3(\vec{s_1}\vec{n})(\vec{s_2}\vec{n}) - \vec{s_1}\vec{s_2} \rangle}{\mu_1 \mu_2} + \Delta_{\SE}, \\
&\Delta_{\SE} = - \frac{2 \sigma}{\pi} \left( \frac{1}{\mu_1} + \frac{1}{\mu_2} \right) \eta \mbox{;}\quad \eta \sim 0.9 \div 1.
\eea
$\Delta_{\SE}$ is the quark self-energy term due to field correlators. Factor $\eta$ is a calculable function of
current quark masses, but is close to 1 when quark masses are small.

Next, we plug in numbers:
\bea
m_u &= 0.005 \;\mbox{GeV}, \,\, m_d = 0.009 \;\mbox{GeV}, \,\, m_s = 0.17 \;\mbox{GeV}, \\
\sigma &= 0.18 \;\mbox{GeV}^2, \\
\alpha_s &= 0.3,
\eea
and taking into account that lowest state is shifted exactly to physical value of meson mass (due to
Gell-Mann-Oakes-Renner relations), we finally get the following chiral shift of reference (quark model)
spectra:

\vspace{0.5 cm}
\noindent pions:
\bea
\nonumber
&\pi(1S) & 0.51 \;\mbox{GeV} \to 0.14 \;\mbox{GeV} &\;(\mbox{exact}) &\\
&\pi(2S) & 1.51 \;\mbox{GeV} \to 1.25 \;\mbox{GeV}&\;(\mbox{exp}:\; 1.3 \;\mbox{GeV}) \\
&\pi(3S) & 2.18 \;\mbox{GeV} \to 1.98 \;\mbox{GeV}&\;(\mbox{exp}:\; 1.8 \;\mbox{GeV})
\eea

\vspace{0.5 cm}
\noindent K mesons:
\bea
\nonumber
&K(1S) & 0.63 \;\mbox{GeV} \to 0.49 \;\mbox{GeV} &\;(\mbox{exact}) &\\
&K(2S) & 1.57 \;\mbox{GeV} \to 1.43 \;\mbox{GeV}&\;(\mbox{exp}:\; 1.46 \;\mbox{GeV})\\
&K(3S) & 2.21 \;\mbox{GeV} \to 2.1 \;\mbox{GeV}&\;(\mbox{exp}:\;1.83 \;\mbox{GeV})
\eea

It can be seen, that masses of radial excitations are shifted by
less than $15 \%$, and the shifts are small for high excitations.
Moreover, one can estimate that $\delta M/M (4S) \simeq  0.05$ for
pions and $\delta M/M (4S) \simeq  0.04$ for K mesons. Also, one
can see that masses of higher excitations and the slope of radial
Regge trajectory differ from the experimental. The reason is that
Hamiltonian~(\ref{eq_Hamiltonian}) does not take into account
effects of string breaking, which are important for highly excited
states, since they have large spatial extent. As was shown in~\cite{Badalian:2002xy}
the inclusion of string breaking effects does not violate the linearity of radial
trajectories, which is in agreement with experimental data.
\footnote{The notion of linear radial Regge trajectories is connected to the
Veneziano-type "linear dual models" \cite{Kataev:1982xu}.}

\newsect
Effective Chiral Lagrangian~(\ref{eq_ECL}) is derived directly from QCD Lagrangian in the framework of Vacuum
Correlators Method. This Lagrangian correctly describes light pseudoscalar mesons, which are massless
in the chiral limit and satisfy Gell-Mann-Oakes-Renner relations when quark masses are nonzero.

Poles of quark model Green's function are shifted by less then $15\%$, and the shift is small for
highly excited states.

The authors are grateful to A.M.~Badalian for valuable comments and discussions.

This work is supported by NSh-1774.2003.2 grant and INTAS-00110 and INTAS-00366 grants.

\end{document}